\documentclass[
    aip,
    apl,
    reprint,
    amsmath,amssymb,
    floatfix
]{revtex4-1}

\usepackage{graphicx}
\usepackage{dcolumn}
\usepackage{bm}
\usepackage{siunitx}
\usepackage{hyperref}
\usepackage[version = 4]{mhchem}

\begin{document}

\title{Phase Formation and Thermal Stability of Superconducting Platinum Silicide Thin Films on Silicon}

\author{Tharanga R. Nanayakkara}
\affiliation{Center for Functional Nanomaterials, Brookhaven National Laboratory, Upton, New York 11973, USA}
\affiliation{Co-design Center for Quantum Advantage, Brookhaven National Laboratory, Upton, New York 11973, USA}

\author{Ananya Chattaraj}
\affiliation{Center for Functional Nanomaterials, Brookhaven National Laboratory, Upton, New York 11973, USA}
\affiliation{Co-design Center for Quantum Advantage, Brookhaven National Laboratory, Upton, New York 11973, USA}

\author{Mingzhao Liu}
\email{mzliu@bnl.gov}
\affiliation{Center for Functional Nanomaterials, Brookhaven National Laboratory, Upton, New York 11973, USA}
\affiliation{Co-design Center for Quantum Advantage, Brookhaven National Laboratory, Upton, New York 11973, USA}

\author{Charles T. Black}
\email{ctblack@bnl.gov}
\affiliation{Center for Functional Nanomaterials, Brookhaven National Laboratory, Upton, New York 11973, USA}
\affiliation{Co-design Center for Quantum Advantage, Brookhaven National Laboratory, Upton, New York 11973, USA}


\date{\today}

\begin{abstract}
Platinum silicide (PtSi) thin films are promising for silicon-based superconducting quantum devices due to their compatibility with CMOS fabrication, air stability, and superconducting transition temperature near 1 K. We report a systematic study of PtSi phase formation, microstructure, and interface quality as a function of annealing temperature and duration, characterizing films using grazing-incidence X-ray diffraction, X-ray reflectivity, and electrical transport measurements. Phase-pure PtSi forms within minutes by rapid thermal processing at 600 °C and is stable under extended annealing, while 30 s anneals across 300–600 °C yield equivalent film quality with consistent microstructure and superconducting properties. X-ray reflectivity reveals that interfacial roughening is an intrinsic consequence of the \ce{Pt2Si}-to-PtSi conversion step rather than a result of elevated temperature or prolonged annealing. These results establish a robust processing window for PtSi formation in silicon-based superconducting device fabrication flows.
\end{abstract}

\maketitle

\section{Introduction}

Platinum silicide (PtSi) is a promising candidate for superconducting quantum devices due to its favorable material properties and compatibility with silicon processing. Formed at relatively low temperatures through well-controlled metal--silicon reactions, PtSi can be integrated with CMOS-based fabrication process flows without excessive thermal budgets. It is chemically stable in air, eliminating the need for encapsulation or passivation during fabrication and storage.

Previous studies~\cite{Hardy1953, Baturina2001, Francheteau2017, Vethaak2021, Nanayakkara2024, Nanayakkara2026} have shown that PtSi films as thin as $\sim$20~nm, or even thinner,~\cite{Oto1994, Yao2024} exhibit superconductivity with a transition temperature near 1~K, and have a high kinetic inductance~\cite{Szypryt2016} and a Ginzburg--Landau coherence length of tens of nanometers~\cite{Oto1994, Nanayakkara2024, Nanayakkara2026}.  These properties are well suited for superconducting applications  requiring compact, high-impedance elements, including microwave resonators, photon detectors, and qubit designs that benefit from large kinetic inductance, such as fluxonium.

In this work, we study the PtSi silicidation process in the context of silicon-based superconducting quantum devices. Our study focuses on the phase formation, microstructure, and interface quality of PtSi films formed on silicon. We characterize the film structural properties using X-ray diffraction and X-ray reflectivity. We determine the film normal-state resistance, residual resistivity ratio, and superconducting transition temperature by electron transport measurements.

\section{PtSi Phase Formation and Stability}

PtSi thin films are typically synthesized by depositing platinum onto crystalline silicon and thermally annealing to induce a solid-state reaction. Platinum deposition has been carried out by physical vapor deposition methods (sputtering or electron-beam  evaporation)~\cite{Naem1988, Pant1992, Larrieu2003} or atomic layer deposition.~\cite{Shi2017}  Annealing is performed in an oxygen-free inert ambient or vacuum,~\cite{Naem1988, Stark2000, Pant1992, Larrieu2003} using either isothermal annealing~\cite{Conforto2001} or rapid thermal processing.~\cite{Naem1988, Pant1992, Larrieu2003}   These conditions generally result in uniform and phase-pure PtSi films, with relatively sharp interfaces to the underlying silicon.

Here, we first examine the PtSi phase composition and microstructure upon rapid thermal annealing at 600~$^\circ$C for extended durations, a temperature well above the reported PtSi formation threshold of $\sim$300~$^\circ$C,\cite{Naem1988, Pant1992, Larrieu2003} to establish a baseline for phase-pure PtSi formation. 
Our synthesis process involved depositing a nominal 10 nm of Pt by sputtering onto silicon substrates that were H-passivated by treatment with buffered oxide etch (BOE) just prior to loading into the vacuum system.

\subsection{Grazing-Incidence X-Ray Diffraction}

\begin{figure}[t]
    \centering
    \includegraphics[width=\columnwidth]{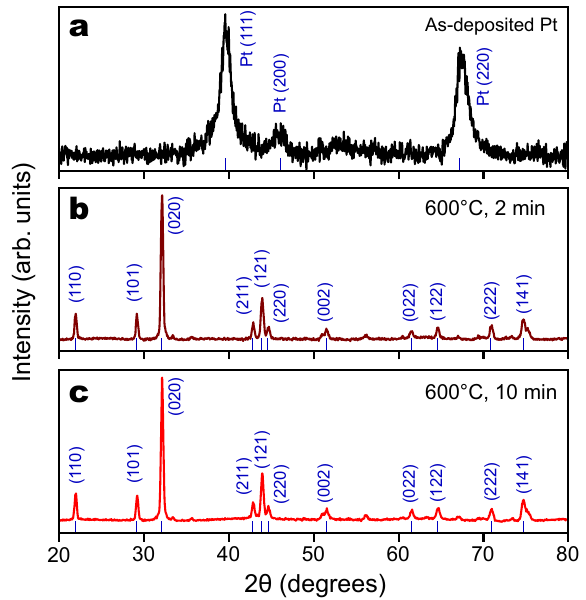}
    \caption{GIXRD patterns of (a) as-deposited Pt film, (b) PtSi after 2~min anneal at 600 °C, and (c) PtSi after 10~min anneal at 600 °C. Reference peak positions and crystallographic indices are indicated in blue.}
    \label{fig:gixrd}
\end{figure}

Grazing-incidence X-ray diffraction (GIXRD) measurements of the as-deposited Pt film show diffraction peaks corresponding exclusively to metallic Pt (Fig.~\ref{fig:gixrd}a).  Annealing at 600~$^\circ$C for 2~min changes the film diffraction pattern to that consistent with single-phase PtSi, with no detectable residual Pt or intermediate \ce{Pt2Si} (Fig.~\ref{fig:gixrd}b). The peak positions and relative intensities are largely unchanged in samples annealed for as long as 10~min (Fig.~\ref{fig:gixrd}c), indicating that the PtSi film remains stable after prolonged 600~$^\circ$C heating. The extended annealing does not induce microstructural coarsening, as determined from analysis of PtSi diffraction peak widths using  Scherrer-type estimates of grain size \cite{Yao2024}. Within our experimental uncertainty, no systematic increase in grain size is observed as the annealing time is increased from 2~min to 10~min (Supporting Information, Fig. S1).

\subsection{Electrical Transport}

\begin{figure}[t]
    \centering
    \includegraphics[width=\columnwidth]{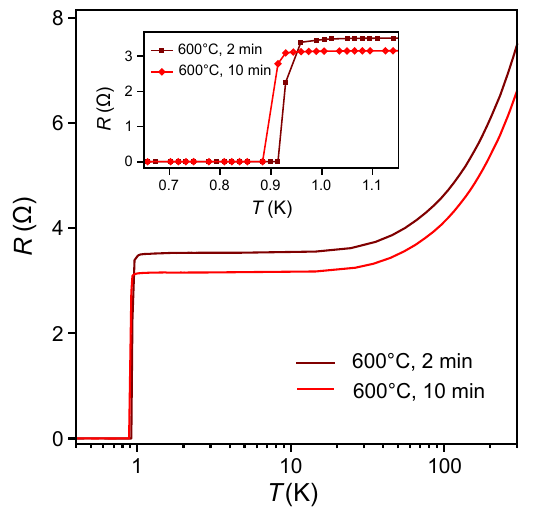}
    \caption{Electrical resistance of PtSi films annealed at 600 °C for (maroon) 2 min and (red) 10 min versus temperature. Inset:  Low-temperature resistance versus temperature shows detail of the superconducting transition. }
    \label{fig:rvt}
\end{figure}

Electrical transport measurements further support that the film properties stabilize shortly after PtSi formation. Films annealed for $2-10$~min exhibit similar room-temperature 4-probe resistance ($\sim$6–7 $\Omega$), residual resistance ratio ($RRR \equiv R_{300\,\mathrm{K}} / R_{2\,\mathrm{K}} = 2.0$), and superconducting transition temperature ($T_c=0.9$ K) (Fig. \ref{fig:rvt}).

\subsection{X-Ray Reflectivity}

\begin{figure}[t]
    \centering
    \includegraphics[width=\columnwidth]{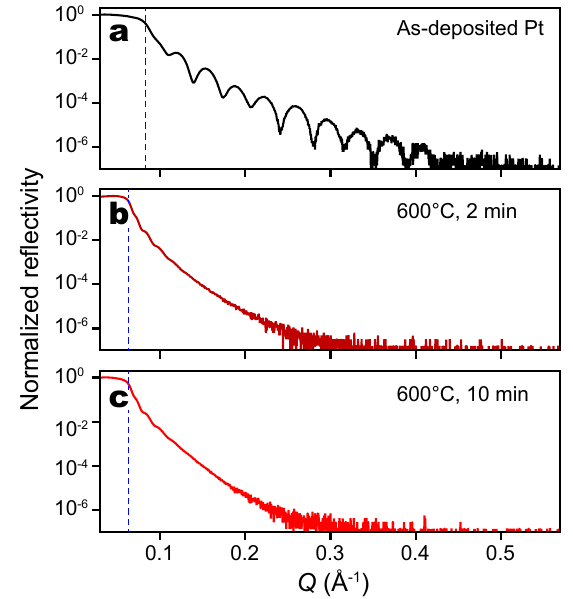}
    \caption{X-ray reflectivity spectra for (a) as-deposited Pt film, (b) PtSi after 2~min anneal, and (c) PtSi after 10~min anneal. $Q=4\pi\sin\theta/\lambda$ is the momentum transfer vector. In each panel, the dashed vertical line denotes the critical edge value $Q_c$, defined as the $Q$ value at the first negative peak of $d^2\log R(Q)/dQ^2$.}
    \label{fig:xrr}
\end{figure}

While XRD and transport measurements establish phase purity and electronic quality, XRR provides independent access to film thickness and PtSi/Si interface roughness. The critical edge of the as-deposited Pt film occurs at $Q_c\approx0.083$~\AA$^{-1}$, corresponding to a density 95\% of bulk Pt ($\rho_\mathrm{Pt, bulk}=$ \qty{21.45}{g.cm^{-3}}). The film shows a series of well-defined Kiessig oscillations extending to $Q = 0.4$~\AA$^{-1}$, with $\Delta Q=0.036$~\AA$^{-1}$, corresponding to a Pt thickness of 17.5 nm (Fig. \ref{fig:xrr}a).  After annealing at 600$^\circ$C, the PtSi films exhibit smaller $Q_c = 0.063$~\AA$^{-1}$ (Fig. \ref{fig:xrr}b, c), corresponding to a density that is 92\% of bulk PtSi ($\rho_\mathrm{PtSi, bulk}=$ \qty{12.28}{g.cm^{-3}}). The Kiessig oscillations become much weaker and disappear by $Q \approx 0.15$~\AA$^{-1}$, with $\Delta Q=0.016$~\AA$^{-1}$ that corresponds to a thickness of 39 nm. The thickness is roughly twice that of the deposited Pt, consistent with volume expansion during the Pt to PtSi conversion. The weakened Kiessig oscillations suggest that the PtSi layer is significantly rougher than Pt, which likely arises from volume expansion and interdiffusion during the Pt-to-PtSi transformation.\cite{Naem1988, Conforto2001}  The XRR spectra do not evolve appreciably with annealing time beyond 2~min (Fig. \ref{fig:xrr}), indicating that interface roughness is established during initial silicidation and does not further degrade with prolonged high-temperature exposure.

Taken together, these results show that PtSi formation at 600~$^\circ$C completes within minutes and is structurally stable under extended annealing times. This processing window  provides a useful reference point for evaluating lower-temperature and shorter-time anneals for use in device fabrication.

\section{Lower-Thermal-Budget Formation of {PtSi} Films}
To evaluate the effects of processing conditions with lower thermal budgets, we examined PtSi formation during 30 s rapid thermal anneals at temperatures ranging between 300–600 °C. This series allows direct comparison with the extended 600 $^\circ$C anneals of Section II, isolating the role of temperature from that of annealing duration.

At 300°C, which lies near the \ce{Pt2Si}-to-PtSi phase boundary under our processing conditions (Section IV), we observe some variability in reaction completion between nominally identical samples, with GIXRD patterns ranging from predominantly PtSi to mixed-phase. This sensitivity is consistent with the steep temperature dependence of the silicidation reaction near the phase boundary, and reinforces that 300°C should be considered a lower bound of the reliable processing window rather than a well-centered operating point.

\begin{figure}[htbp]
    \centering
    \includegraphics[width=\columnwidth]{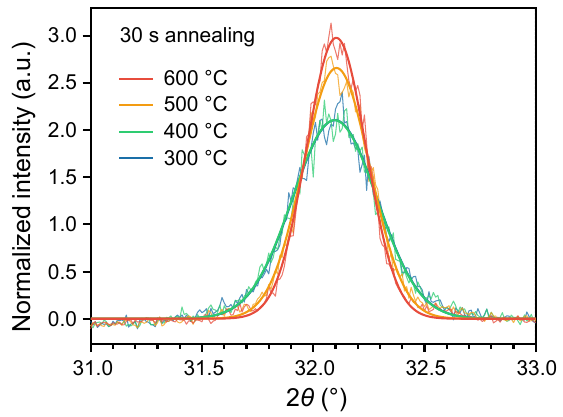}
    \caption{Normalized PtSi (020) GIXRD diffraction peak for films annealed for 30 s at temperatures from 300 to 600 $^\circ$C. Solid lines are Gaussian fits to the measured data. The decreasing peak width with increasing annealing temperature indicates increasing crystallite size.}
    \label{fig:ptsi_020_tempseries}
\end{figure}

Crystallite size estimates extracted from Scherrer analysis of the PtSi (110) and (020) diffraction peak widths indicate a modest but systematic increase with annealing temperature across the 300–600 °C range (Fig.~\ref{fig:ptsi_020_tempseries}). The coherent domain size increases from $\sim20 \pm 1$ nm at 300–400 °C to $\sim24  \pm 1$ nm at 450 $^\circ$C, reaching $\sim27 \pm 1$ nm at $600^\circ$C. Notably, this larger size is similar to that observed for the $600^\circ$C, 10 min anneal, indicating that crystallite growth saturates rapidly upon PtSi formation and is insensitive to annealing duration once silicidation is complete.

\begin{figure}[t]
    \centering
    \includegraphics[width=\columnwidth]{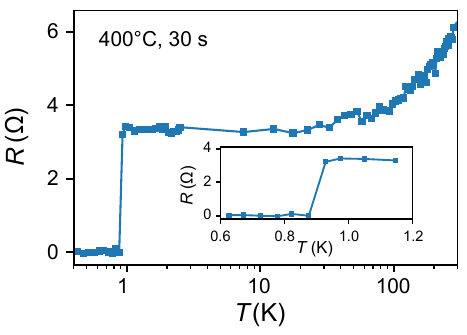}
    \caption{Resistance versus temperature from 300~K to 500 mK for a PtSi film formed by annealing at 400°C for 30 s. Inset highlights the low-temperature superconducting transition. }
    \label{fig:rvt_low_temp_samples}
\end{figure}

Electrical transport measurements further support the consistency of film microstructure across the annealing temperature range. Films annealed at $400^\circ$C and 450$^\circ$C for 30 s exhibit comparable room-temperature resistance ($\sim$6  $\Omega$) and $T_c$ (0.9 K) (e.g., 400 °C annealing film in Fig. \ref{fig:rvt_low_temp_samples}), and these are similar to those measured for the longer-duration $600 ^\circ$C anneals (Section II). The RRR ($\sim1.5$) is lower than that measured for films annealed at 600°C (Section II), consistent with the smaller film crystallite size and more grain boundaries, which provide additional scattering channels. That $T_c$ remains unchanged indicates that the superconducting condensate is not significantly affected by the modest variation in grain size.

X-ray reflectivity measurements give further information about the film roughness and interface quality. Throughout the entire range of 300–600 °C annealing temperatures, all spectra for the short-time annealed samples have $Q_c = 0.063$~\AA$^{-1}$ that is identical to those obtained after long-duration $600^\circ$C silicidation, with very similar Kiessig oscillations (Supporting Information, Fig. S3). This indicates that the PtSi/Si interface quality is established during initial stages and does not evolve with longer thermal processing. 

These results show that 30 s anneals produce PtSi films with stable microstructure and reproducible superconducting properties across a wide processing window spanning $300-600^\circ$C. The robustness of film quality across this range provides flexibility for integration into diverse device fabrication flows.

\section{Sequential \NoCaseChange{PtSi} Silicidation Kinetics and Phase Evolution}

The PtSi reaction pathway is known to be a two-step, diffusion-limited process, with Pt first diffusing into crystalline Si to form an intermediate \ce{Pt2Si} phase, typically initiating at temperatures near $250^\circ$C.\cite{Naem1988, Pant1992, Larrieu2003}  Once the Pt supply is consumed, a second reaction proceeds in which Si diffuses into \ce{Pt2Si}, yielding the thermodynamically stable PtSi phase at temperatures above $\sim$$300^\circ$C.\cite{Naem1988, Pant1992, Larrieu2003} This sequential synthesis mechanism has been confirmed in multiple studies using in situ ellipsometry, Rutherford backscattering spectroscopy, and X-ray photoelectron spectroscopy.\cite{Stark2000, Pant1992, Larrieu2003} The consistency of these findings across both ramped and isothermal annealing conditions has provided a  quantitative basis for modeling silicide kinetics in device-relevant regimes.\cite{Stark2000, Conforto2001}

To probe these kinetics and the evolution of film interface microstructure, we performed a series of single-step anneals at temperatures spanning the two reaction thresholds, and a sequential two-step anneal to isolate the intermediate \ce{Pt2Si} phase before driving the second reaction to completion.

\begin{figure}[htbp]
    \centering
    \includegraphics[width=\columnwidth]{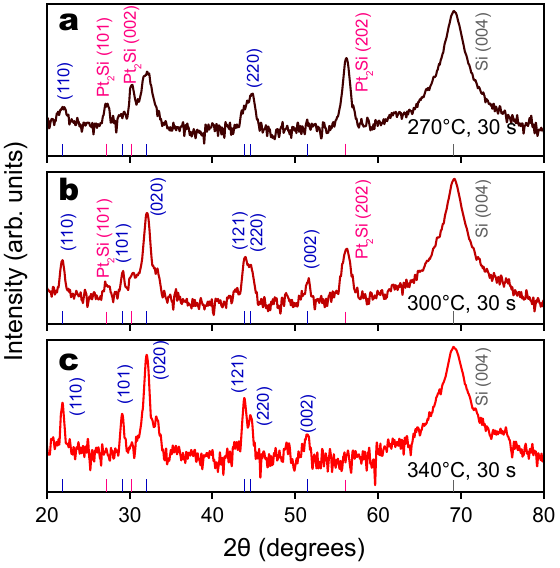}
    \caption{X-ray diffraction $\theta$--$2\theta$ scans of Pt/Si films annealed for 30 s at (a) 270 $^\circ$C, (b) 300 $^\circ$C, and (c) 340 $^\circ$C. The diffraction patterns show the progression from \ce{Pt2Si} to mixed \ce{Pt2Si}/PtSi and finally single-phase PtSi with increasing annealing temperature.}
    \label{fig:ptsi_lowT_annealing_XRD}
\end{figure}

XRD $\theta-2\theta$ scans from the single-step anneal series reveal the phase boundaries (Fig. \ref{fig:ptsi_lowT_annealing_XRD} and Supporting Information, Fig. S4). The $260^\circ$C anneal yields diffraction peaks that are very similar to as-deposited metallic Pt, indicating that silicidation has not yet initiated (Supporting Information, Fig. S4). At $270^\circ$C, the diffraction pattern changes to that of \ce{Pt2Si}, with no detectable residual Pt or PtSi (Fig. \ref{fig:ptsi_lowT_annealing_XRD}a), confirming that the first reaction step is complete within our 30 s anneal time at this temperature. At 300°C (Fig. \ref{fig:ptsi_lowT_annealing_XRD}b), the pattern shows both PtSi and \ce{Pt2Si}, consistent with the onset of the second reaction step.  Above $300^\circ$C (Fig. \ref{fig:ptsi_lowT_annealing_XRD}c), the pattern is consistent with pure PtSi. These observations establish the phase boundaries of the sequential reaction under our rapid thermal processing conditions.

\begin{figure}[htbp]
    \centering
    \includegraphics[width=\columnwidth]{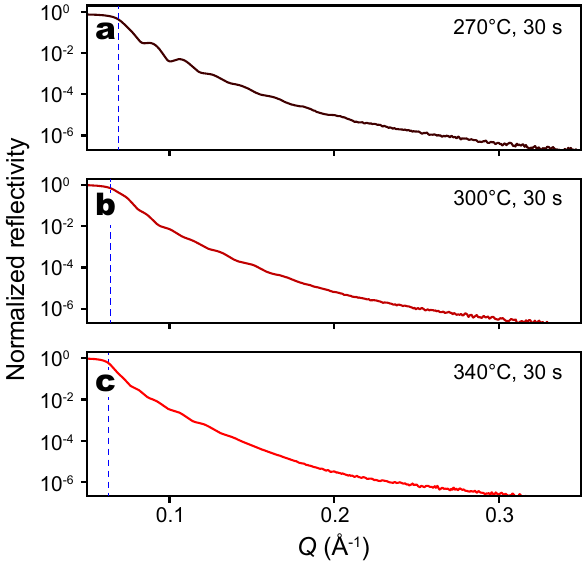}
    \caption{XRR measurements of Pt/Si films annealed for 30 s at (a) 270 $^\circ$C, (b) 300 $^\circ$C, and (c) 340 $^\circ$C. The \ce{Pt2Si} film formed at 270 $^\circ$C exhibits stronger Kiessig oscillations than films containing PtSi, consistent with increased interface roughness upon PtSi formation.}
    \label{fig:ptsi_lowT_annealing_XRR}
\end{figure}

XRR measurements across this single-step series reveal an asymmetry in interface evolution between the two reaction steps (Fig. \ref{fig:ptsi_lowT_annealing_XRR}). The \ce{Pt2Si} film formed at $270^\circ$C exhibits well-defined Kiessig oscillations extending to $Q = 0.3$ \AA$^{-1}$ (Fig. \ref{fig:ptsi_lowT_annealing_XRR}a), with $Q_c = 0.069$~\AA$^{-1}$, which is clearly larger than the value observed for PtSi ($\sim 0.063$~\AA$^{-1}$) and better matched to \ce{Pt2Si}. In contrast, the silicide films formed at $300^\circ$C and above show $Q_c$ similar to bulk PtSi, with weaker Kiessig oscillations (Fig. \ref{fig:ptsi_lowT_annealing_XRR}b, c).  The increase in roughness upon conversion from \ce{Pt2Si} to PtSi suggests that the interfacial disruption associated with silicidation occurs predominantly during the second reaction step, when silicon diffuses into the \ce{Pt2Si} layer, rather than during the initial Pt-into-Si reaction.\cite{Conforto2001}

\begin{figure}[htbp]
    \centering
    \includegraphics[width=\columnwidth]{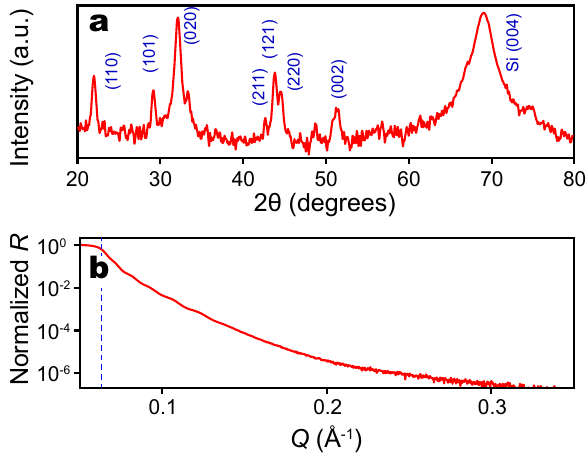}
    \caption{(a) XRD $\theta$--$2\theta$ scan and (b) XRR measurement of a Pt/Si film formed by sequential annealing (280 $^\circ$C, 30 s followed by 360 $^\circ$C, 30 s). The XRD pattern is consistent with single-phase PtSi, while the XRR profile exhibits weak Kiessig oscillations comparable to those of single-step PtSi films.}
    \label{fig:ptsi_2step_annealing_XRD_XRR}
\end{figure}

To examine whether a two-step pathway can influence the final film microstructure, we prepared a sequentially annealed sample consisting of a first anneal at $280^\circ$C for 30 s to form \ce{Pt2Si}, followed by a second anneal at $360^\circ$C for 30 s to drive conversion to PtSi. XRD $\theta-2\theta$ scans confirm that the final film is single-phase PtSi (Fig. \ref{fig:ptsi_2step_annealing_XRD_XRR}a), consistent with complete reaction through the sequential pathway. XRR measurements of the two-step sample yield a $Q_c=0.063$~\AA$^{-1}$ with weak Kiessig oscillation (Fig. \ref{fig:ptsi_2step_annealing_XRD_XRR}b), closely resembling those of the single-step PtSi films formed at comparable temperatures. This indicates that routing the reaction through the \ce{Pt2Si} intermediate does not meaningfully alter the final interface quality 
--- the roughening incurred during the second reaction step appears to be an intrinsic consequence of the \ce{Pt2Si}-to-PtSi conversion, independent of how the intermediate is formed.\cite{Larrieu2003}

Combining the single-step and two-step results, we identify the second reaction step (Si diffusion into \ce{Pt2Si}) as the primary source of interfacial roughening. This mechanistic insight suggests that strategies for improving PtSi interface quality should focus on controlling the second reaction step.

\section{Conclusions}

We have characterized the phase formation, microstructure, and interface quality of PtSi thin films formed by Pt silicidation on silicon by rapid thermal annealing over a wide range of processing conditions. PtSi films formed at temperatures near $300^\circ$C and above, and for durations as short as 30 s, exhibit stable microstructure and retain superconducting properties comparable to those formed under more aggressive thermal treatments. XRR measurements reveal that interfacial roughening is an intrinsic consequence of the \ce{Pt2Si}-to-PtSi conversion step, rather than a result of extended or elevated-temperature annealing. These findings establish a robust processing window for PtSi formation, supporting its integration into silicon-based superconducting device fabrication flows across a range of thermal budget constraints.

\begin{acknowledgments}
This material is based upon work supported by the U.S. Department of Energy, Office of Science, National Quantum Information Science Research Centers, Co-design Center for Quantum Advantage under contract number DE-SC0012704. This research used Materials Synthesis \& Characterization facility of the Center for Functional Nanomaterials (CFN), which is a U.S. Department of Energy Office of Science User Facility, at Brookhaven National Laboratory under Contract No. DE-SC0012704. 
\end{acknowledgments}

\section*{References}

\bibliography{apssamp}

\end{document}